\documentclass[11pt,twoside]{article}
\usepackage{./asp2014}

\aspSuppressVolSlug
\resetcounters

\begin{document}

\title{Be Stars as Seen Through Telescopes in Survey Mode (II)} 
\author{Th.~Rivinius,$^{1}$
  C.~Martayan,$^{1}$ and D.~Baade$^{2}$
\affil{$^{1}$ESO -- European Organisation for Astronomical Research in the
  Southern Hemisphere, Chile; email: {\tt triviniu@eso.org}}
\affil{$^{1}$ESO -- European Organisation for Astronomical Research in the
  Southern Hemisphere, Germany}
}

\paperauthor{Th.~Rivinius}{triviniu@eso.org}{}{ESO - European Organisation for Astronomical Research in the Southern Hemisphere}{}{Vitacura}{}{}{Chile}
\paperauthor{D.~Baade}{dbaade@eso.org}{}{ESO - European Organisation for Astronomical Research in the Southern Hemisphere}{}{Garching}{}{}{Germany}
\paperauthor{S.~\v{S}tefl}{}{}{ESO/ALMA - The Atacama Large Millimeter/Submillimeter Array}{}{Santiago}{}{}{Chile}
\paperauthor{R.H.D.~Townsend}{}{}{Department of Astronomy, Univ.\ of Wisconsin-Madison}{}{Madison/WI}{}{}{USA}
\paperauthor{A.C.~Carciofi}{}{}{Instituto de Astronomia, Geof\'isica e Ci\^encias Atmosf\'ericas}{}{Sao Paulo}{}{}{Brazil}

\begin{abstract}
The first half of the review dedicated to survey works on Be stars (Baade,
Martayan, and Rivinius, this vol.)  put emphasis on what we can learn from
surveys about Be stars as a part of an environment, such as Be stars in
binaries, Be stars in different metalicities, or Be stars as part of a star
forming and then co-evolving group. This second half will rather concentrate
on the information that more focused surveys can give on a Be star, understood
as an individual object, and in this way attempts to bridge the gap between
highly detailed single star studies, and necessarily broad survey and catalog
work.
\end{abstract}

\section{Introduction}
Surveys that have investigated Be stars in a more focused way usually include
no more than a few dozen objects. Properties studied in such surveys include
the stellar rotation of Be stars, their chemical surface abundances, the
pulsational and magnetic properties, and the life cycles and evolution of the
circumstellar disk.

The common theme between most of these topics is the question how,
actually, does a Be star form its disk, and in particular how is the mass
ejected with sufficient angular momentum to remain in orbit around the star?
The answer is not yet given, but survey results have the potential to narrow
down the candidate list to just a few processes, that can then possibly be
tested with reasonable effort by more detailed, single star studies.

\section{Rotation}
The question of how rapidly do Be stars rotate was considered to be settled a
while ago. Before about the year 2000, the consensus held that Be stars as a
class rotate at about 80\% of their critical value (i.e. at which material
would escape from the equator without an additional lift-off mechanism
required), but rarely above that or even at critical value. Interferometric
observations of Achernar shook this consensus. \citet{2003A&A...407L..47D}
reported that the star was rotationally flattened so much that it could only
be explained by 100\% critical rotation (or even faster). Soon afterwards,
\citet{2004MNRAS.350..189T} explained why the 80\% critical value could well
be the result of an observational selection effect, namely that the most
rapidly rotating parts of the star, the equator, would become so dim due to
gravity darkening that it would no longer contribute to the line width. This
would mean that the observed $v \sin i$ becomes degenerate in terms of the
true rotation, so that all stars rotating at and above 80\% would be measured
to rotate precisely at, but not with more than 80\%.

Since then, several studies have aimed to determine the actual Be star
rotation rate. Advances in interferometric instrumentation have allowed to
re-measure Achernar with a precision that was impossible ten years ago
\citep{2014A&A...569A..10D}. As a result, Achernar is now the Be star with the
best known stellar parameters, and probably so by a fair margin. In terms of
rotation, it was found that the the original result had to be explained by
contamination due to a weak circumstellar disk. The new observations, taken in
a diskless phase, revealed a rotation rate of 88\% critical, which translates
to 84\% of the velocity needed to lift a particle into orbit just above the
actual equator.

Interferometry as well served to resolve one of the largest ambiguities in
measuring rotation. \citet{2012A&A...538A.110M} determined the inclination
angle for a number of stars, and with this could calculate the true rotational
rates with better accuracy than by standard techniques. This and a number of
other survey works \citep[see Sect.~3.1][for a review]{2013A&ARv..21...69R}
confirmed that Be stars can rotate significantly sub-critical. However, the
rotational rates are not distributed narrowly around 80\%, but rather spread
between about 75\% and 100\%. In other words, once a B star rotates above that
threshold, it can become a Be star. This threshold does not depend on the
spectral subtype. 

From a different perspective, one can ask as well whether there is a threshold
above which a B star must become a Be star, i.e.\ no more non-emission B stars
are observed. According to \citet{2010ApJ...722..605H} this is the case, but
here it {\em does} depend on the spectral subtype: While all early type B
stars above 75\% critical rotation are also Be stars, the limit above which
only Be stars exists increases to 90\% in late type B stars. For the formation
of Be stars, this means that there are either several processes that are
differently weighted against each other at the different spectral subtypes, or
in case that there is only one such process it must decrease in efficiency
from early to late B subtypes.

\section{Chemical Surface Abundances}

Closely related to the question of rotation of Be stars is the one of the
chemical abundances at the surfaces of Be star, potentially modulated due to
rotational mixing. For slowly and intermediately fast rotating stars,
rotational mixing is well understood and observationally calibrated. However,
the typical rotation rates used for calibrating the models are still below the
typical rotation rates of Be stars. If this scheme of rotational mixing is
extrapolated into the Be star regime, one finds that a significant fingerprint
of mixing should arise in a relatively short time, so much that it should be
large enough to be detectable even in the very shallow lines that make rapidly
rotating stars difficult to analyze, and very definitively in pole-on Be stars.

However, only few studies were available until recently. Single stars and
small samples were investigated by \citet{1986ESASP.263..377H},
\citet{2005A&A...438..265L}, and \citet{2011IAUS..272..101P}, who all found
negative result (i.e.\ no significant rotational mixing modulation or other
enrichment), and by \citet{2005A&A...442..263V} and
\citet{2011NewA...16..307L}, who find a chemical enrichment pattern, but due
to the particular patterns obserevd favor binary interaction as explanation,
rather than rotational mixing. The only survey type study was undertaken by
\citet{2011A&A...536A..65D}, who did not find strong Nitrogen enrichment in 30
Be stars of the LMC and SMC.

In other words, the observed surface abundances of Be stars are inconsistent
with the values predicted by rotational mixing for typical Be star rotational
velocities.  One can suggest a number of hypotheses to explain such a result:
First, Be stars could rotate much slower than thought, but this is not very
likely in light of the previous section. Second, they could become rapid
rotators only very shortly before or contemporary with becoming Be stars, and
Be stars do not stay Be stars for very long; but this is hardly consistent
with the incidence and statistics of Be stars. Third, somewhere above the
limit at which mixing models are calibrated things go wrong, but this would
require a new ingredient in the theory of stellar constitution, such as a
shell/layer inside the star that blocks mixing, but arises only at high
rotation rates.

However, the abundance analysis of Be stars comes with a particular set of
problems, partly due to the rapid rotation that forbids treating Be star line
formation with a single value of temperature and gravity, and partly due to
the presence of line emission and the scattered continuum from the disk. A
further confirmation of having primordial (meaning here: as the star was
formed) abundances in Be stars is certainly required before overturning the
established theory of rotational mixing, but neither can the studies pointing
towards the need for such a revision be ignored.

\section{Pulsation}

Be stars, as a class, are pulsating stars. With ground-based observations the
proof for pulsational variability could be delivered only for a limited set of
objects, mostly early type stars with relatively high amplitudes \citep[][and
  references therein]{2003A&A...411..229R}. Nowadays, several years into the
era of space-based time-series photometry, however, the question whether Be
stars are pulsators or not is settled. Of more than thirty Be stars that were
observed by asteroseismology satellites so far, every single one was found to
be multiperiodic \citep[see Sect.~3.2 of][for an
  overview]{2013A&ARv..21...69R}. One has to stress that this does not mean
that every single of those periods is due to pulsation, quite to the contrary
do the data indicate that there is additional variability to the pulsational
one, which we do not fully understand yet. Strictly speaking, there is as well
a lack in understanding the pulsation: current theory of pulsational
excitation does not cover stars rotating as rapidly as Be stars.

The relevant question for Be stars is not so much whether they pulsate or not,
but if and how the pulsation is linked to their nature as Be stars, in
particular since pulsation in the upper main sequence seems not to be an
exception, but rather the rule. \citet{2011MNRAS.413.2403B} report an
incidence of 30\% pulsating B stars. However, they discard Be stars from their
list, so if Be stars are added in the fraction is closer to one half, which is
a lower limit. Can pulsation contribute to the angular momentum transfer into
the disk? Here the picture is much less clear. There are multiperiodic stars
in which co-added amplitude maxima (beating) trigger outbursts
\citep[$\mu$\,Cen, possibly 28\,CMa and $\eta$\,Cen:
  see][]{1998BeSN...33...15R,2000ASPC..214..232T,2003A&A...411..229R}. As well
some stars show either short-term amplitude change that is correlated with the
mass-ejection (some asteroseismology targets, including, for instance,
Achernar; \citealt{2011MNRAS.411..162G}), or a pulsational phase drift during
the mass-ejection episodes
\citep[$\omega$\,CMa:][]{2003A&A...411..167S}. However, how all these
observations could possibly be merged into a unified picture of a
pulsationally modulated and rotationally supported mass ejection is completely
unclear.

Kee et al. (this vol.) have explored the potential efficiency of angular
momentum transfer and found that only modes with both retrograde phase- and
group-velocity would be incapable of creating and sustaining a Be star
disk. Modes with both prograde velocities would be most efficient, but as well
more exotic modes with retrograde phase and prograde group velocity, or
vice-versa, could work to make a Be star disk. The question which of these
modes exist in Be stars is open. While spectroscopic modeling is fairly robust
and favors retrograde/retrograde modes (which would not work to form a disk),
asteroseismic modeling favors prograde/prograde modes (but is outside its
proven validity regime). So possibly the solution lies indeed in one of the
``mixed'' mode types.

\section{Magnetic Fields}

Just a few years ago, magnetic fields in rapidly rotating, early type stars
were considered to be pretty much out of observational reach. Similar as space
based asteroseismology was a game-changer for the pulsation question, with the
MiMeS survey completed (Magnetism in Massive Stars, Wade et el., this vol.)
the picture has drastically changed for magnetism of early type stars. For
completeness we note that the second extensive survey on the matter (BOB:
B-fields in OB-Stars) is not yet complete, and does not target Be stars
\citep{2015IAUS..307..342M}.

In a simple picture, magnetic fields would elegantly provide the means for the
angular momentum transport into the circumstellar environment, and indeed
initially a few Be stars were reported as magnetic. However, these 
works \citep{2003A&A...409..275N,2007AN....328.1133H,2009AN....330..708H}
reported fields close enough to the detection limit to warrant confirming
observations, in particular when FORS detections came under more general
criticism \citep{2012A&A...538A.129B}. For the stars for which such
confirmation was sought, the result was negative ($\omega$\,Ori:
\citealt{2012MNRAS.426.2738N}, $\chi$\,Oph: \citealt{2009MNRAS.398.1505S}, and
$\mu$\,Cen: Wade, priv.\ comm.).

The MiMeS survey took it a step further by observing about 85 Be stars and
analyzing the results in a firm statistical framework (Wade et el., this
vol.). The result is that Be stars certainly do not possess large scale,
i.e.\ of low multipole order, magnetic fields of any strength above a few
hundred Gauss, and quite possibly not at all. 

However, it should be noted that the presence of a magnetic field and rapid
rotation are not mutually exclusive, but in such stars the circumstellar
environment takes the form of a magnetosphere that is governed by the magnetic
field, rather than a Keplerian disk as in Be stars.

This means that the average Be star is actually less magnetic than the average
non-emission B star (of which about 5 to 10\% possess a kG large scale
magnetic field, see Wade et al., this vol.), and although the MiMeS result
does not entirely rule out small-scale fields, such as magnetic loops, such
fields have not been observed in any early type star yet. The detectability of
such fields depends on several assumptions, so it will always be possible to
argue for a specific geometry to remain undetected. Notwithstanding, with
current capabilities, including MiMeS, the detection could already have been
possible under certain circumstances, as demonstrated by
\citet{2013A&A...554A..93K}. Rather, such small scale fields and are either
merely hypothesized to save the magnetic ejection model that gave rise to the
search for magnetic fields in the first place, or are hinted at only by
indirect evidence, e.g.\ from the X-Ray regime \citep[see Sect.~3.3
  of][]{2013A&ARv..21...69R}.

\section{Disk Properties}

Concerning Be star disks, most survey type work has been done with photometry
obtained during campaigns such as OGLE or MACHO. The currently published
results concentrate mostly on the phenomenological properties of the light
curves on medium and long time scales, i.e. those governed by the built-up
and decay of the disk. However, during this meeting a number of interesting
works have been presented that take it a step further, namely into determining
physical properties of the disks from the shape and amplitude of the observed
variations.

From the theoretical side, the dynamical viscous disk model has achieved a
stunning success in the past decade. It should be kept in mind that any
substantiated criticism to this model is possible not despite, but only
because its success in the quantitative modeling of the behavior of Be star
disks. The model has, for instance, been applied to the the well observed disk
formation and decay phases of $\omega$\,CMa \citep{2012ApJ...744L..15C}. The
result was somewhat surprising, although the decay of the light curve could be
modeled quite well, it required a fairly high value of the turbulent viscosity
parameter, namely $\alpha=1$. Further works since then have confirmed that
result for $\omega$\,CMa (see Ghoreyshi et al., this vol.).

Since the method applied for $\omega$\,CMa relies on photometric data alone,
it can be applied to light curves observed in surveys. Preliminary results,
based on the work by R\'imulo et al. (this vol.\ and priv. comm.) indicate
that a high value of $\alpha$ is the norm for Be star disks. Since $\alpha$
parameterizes the turbulent speed in relation to the speed of sound, a value
of unity is normally considered a natural upper limit, and even that is not
usually observed: Observational determinations of the viscosity parameter in
gaseous disks, e.g.\ of cataclysmic variables, usually derive values one order
of magnitude or more lower.

One question is whether the assumption of using the $\alpha$ prescription is
justified. If a non-viscous process, like radiative ablation, would contribute
to the disk decay this may mimic a high value of $\alpha$. However, results by
Ghoreyshi et al.\ and R\'imulo et al.\ (both op. cit.) indicate that also the
build-up of the disk has to be modeled with a high value of $\alpha$.  If
$\alpha$ is indeed due to turbulent viscosity, there might be a mechanism that
drives the value to its natural limit. The contribution by Fung (this vol.)
has shown the interesting prospect of increased turbulence at the inner edge
of the disk, induced by the radiation pressure of the central star.

On the other hand, some Be stars, including $\omega$\,CMa show an underlying,
very long-term secular trend in the light curve. Every outburst in the last 40
years returned to a somewhat lower base value, and in quiescent times a
downward slope could be observed directly. This type behavior is as well seen
in OGLE data for some stars.  Such a slow trend is hard to explain with a high
$\alpha$, unless one assumes a tuned mass-loss behavior with the same
properties, i.e.\ a long term decrease with overriding outbursts. In turn,
attempts to model that behavior with a value of $\alpha$ that is not constant,
either in time or over the disk (suggested e.g.\ by Kurf\"urst et al., this
vol.) have not been successful, either (Carciofi, priv. comm.). The source and
properties of the viscosity in the otherwise very successful viscous disk
model remain puzzling.

\section{Some Conclusions}
Summarizing what has been learned from surveys, but as well what new
questions were opened:

\begin{itemize}
\item As a group, Be stars rotate rapidly (>75\%), including some even at the
  critical limit.
\item Although the rapid rotation is the most important single factor in
  forming a Be star, it is not sufficient to explain Be stars without further
  mechanism(s) acting.
\item The chemical surface abundance pattern observed in Be stars seems
  inconsistent with the current theory of rotational mixing in rapidly
  rotating stars.
\item Be stars do not possess low-order, large scale magnetic fields.
\item Be stars are pulsating stars, with most (possibly all) Be stars
  pulsating in SPB-like modes. Some Be stars are $\beta$\,Cephei pulsators,
  too \citep[$\pi$\,Aqr, for instance:][]{2005ASPC..337..294P}.
\item The precise nature and excitation of the pulsation modes are
  unclear.
\item In some stars, the pulsation is clearly linked to the disk formation.
\item The nature of this link, or links, is unclear.
\item The disk, once formed, is governed by viscous processes.
\item The viscosity parameter is surprisingly high, close to a natural limit
  of unity. Whether it is constant across the disk and among Be stars as a
  group is unclear.
\end{itemize}


\bibliography{rev}  

\end{document}